\documentclass{emulateapj}

\newcommand{\Msol}{\mbox{M$_\odot$}}

\newcommand{\joz}{\mbox{$J=1 \rightarrow 0$}}
\newcommand{\jto}{\mbox{$J=2 \rightarrow 1$}}

\newcommand{\omm}{\mbox{$\lambda=1.3$ mm}}
\newcommand{\tmm}{\mbox{$\lambda=2.7$ mm}}
\newcommand{\klambda}{\mbox{k$\lambda$}}

\newcommand{\cojto}{$\textrm{CO}~J=2\rightarrow1$}

\slugcomment{Accepted for publication in \apj}

\shorttitle{Grain Growth and Density Distribution in the Youngest Protostellar Systems}
\shortauthors{Kwon et al.}

\begin{document}

\title{Grain Growth and Density Distribution of the 
Youngest Protostellar Systems}

\author{Woojin Kwon\altaffilmark{1}, Leslie W. Looney\altaffilmark{1}, 
Lee G. Mundy\altaffilmark{2}, Hsin-Fang Chiang\altaffilmark{1}, and
Athol J. Kemball\altaffilmark{3}}
\altaffiltext{1}{Department of Astronomy, University of Illinois, 
1002 West Green Street, Urbana, IL 61801; 
wkwon@illinois.edu, lwl@illinois.edu, hchiang2@illinois.edu}
\altaffiltext{2}{Department of Astronomy, University of Maryland, College Park, 
MD 20742; lgm@astro.umd.edu}
\altaffiltext{3}{National Center for Supercomputing Applications, 
University of Illinois, 1205 W. Clark Street, Urbana, IL 61801;
akemball@illinois.edu} 

\begin{abstract}
We present dust opacity spectral indexes ($\beta$) of 
the youngest protostellar systems (so-called Class 0 sources), 
L1448 IRS 2, L1448 IRS 3, and L1157, obtained between
the \omm\ and 2.7 mm continua, using the Combined Array for Research
in Millimeter-wave Astronomy (CARMA).  The unprecedented compact
configuration and image fidelity of CARMA allow a better detection of
the dust continuum emission from Class 0 sources, with a less serious
missing flux problem normally associated with interferometry.  
Through visibility-modeling at both \omm\ and 2.7 mm simultaneously, as
well as image- and visibility-comparison, we show that $\beta$ of
the three Class 0 sources are around or smaller than 1, indicating
that dust grains have already significantly grown at the Class
0 stage. In addition, we find a radial dependence of $\beta$, which
implies faster grain growth in the denser central regions and/or dust
segregation.  Density distributions of the Class 0 sources are also
addressed by visibility-modeling.
\end{abstract}


\section{Introduction}

Although dust grains are only about one hundredth of the interstellar
medium by mass, they play crucial roles for star formation, planet
formation, and furthermore the origin of life. They are essential
places to form and store molecules, and they are the main ingredient
to form terrestrial planets, as well as playing a role in the heating
and cooling mechanisms during star and planet formation.

The dust opacity\footnote{
Dust ``emissivity'' has also been used 
in literatures from the viewpoint of dust thermal ``emission''.
}
spectral index ($\beta$) is related to dust properties.  It
depends on dust grain sizes, compositions, and shapes
\citep[e.g.,][]{pollack1994,draine2006}.  In particular, it is
largely sensitive to grain sizes; larger grains give smaller $\beta$
\citep[e.g.,][]{draine2006}.  Many observational studies at infrared
and millimeter wavelengths toward T Tauri circumstellar disks have
reported smaller values of $\beta$ ($\sim 1.0$) \citep[e.g.,][]{andrews2007}
compared to that of the interstellar medium ($\sim 1.7$)
\citep[][]{finkbeiner1999,li2001}.  In the sense that dust grains
may develop terrestrial planets, it is very encouraging to
see signatures of larger dust grains in T Tauri disks, evolved young
stellar objects (YSOs), compared to grains in the interstellar
medium.

However, it is not clear when the dust grain growth responsible for the
opacity spectral index $\beta \sim 1$ mainly occurs.
For example, while \citet{andrews2005} reported grain growth along
the YSO evolution from Class I to Class II, using spectral energy
distributions over \omm\ and submillimeter data, \citet{natta2007ppv}
did not find such a tendency (a systematic variation of $\beta$). 
To distinguish when dust grains
mainly grow up to the sizes for $\beta \sim 1$, Class 0 YSOs are the best targets to examine.  
Class 0 YSOs are at the starting point
of low-mass star formation and they are well defined. They have
more massive envelopes than or comparably massive envelopes to their 
central compact objects \citep[e.g.,][]{andre1993}.  
They are also characterized with well-developed
bipolar outflows.  Earlier stages such as starless cores might be
another good target but they are hardly confined. Their physical
conditions including age have a much larger scatter than Class 0 sources.
In addition, they are not all expected to form stars.

In fact, no definitive answer has been given to the opacity spectral
index $\beta$ of Class 0 sources so far.  It is another reason that
this study is needed beyond the grain growth point of view.
There are some previous studies about the flux density spectral
indexes of Class 0 sources, which are related to the dust opacity
spectral indexes, although they have not focused on dust properties
\citep[e.g.,][]{hogerheijde2000,shirley2000}.  However, these studies
used submillimeter to 1.3 mm wavelengths, which is near the range of
peak intensities at envelope temperatures ($\sim 30$ K), so the
Rayleigh-Jeans approximation is invalid.  In that case, the estimate
of $\beta$ is sensitive to the envelope temperature, which causes
relatively large uncertainties in the $\beta$ estimate.
In addition, optical thickness can cause another uncertainty, since
Class 0 YSO envelopes can be optically thick at submillimeter
wavelengths.  
On the other hand, \citet{harvey2003} obtained $\beta \sim 0.8$
toward the Class 0 YSO B335 using $\lambda=1.2$ mm and 3 mm
interferometric data, while carrying out modeling to test density
distribution models of star formation.  However, they did not have
a good data set with comparable {\it uv} coverage at both wavelengths
to discuss the $\beta$ in detail.  In other words, there are no reliable
$\beta$ estimates of Class 0 YSOs. As a result, many studies to
estimate masses from spectral energy distributions (SEDs) and/or
to constrain density distributions have assumed $\beta \sim 1$
\citep[e.g.,][]{looney2003} or considered a possible range of $\beta$
\citep[e.g., $\beta = 1-2$,][]{chandler2000}.

Radio interferometry at millimeter wavelengths is the best means
to investigate the $\beta$ of Class 0 YSOs.  As mentioned, optical
thickness and dust temperature issues cause large uncertainties at
shorter wavelengths. On the other hand, contamination of non-thermal
continuum increases with wavelength, so it is not negligible
at longer centimeter wavelengths.  In addition, considering envelope
sizes of Class 0 YSOs and their environments (normally they are
within extended molecular clouds), single dish observations are not
appropriate due to their lack of angular resolution and the
contamination of molecular clouds.  In contrast, interferometers
provide high angular resolution and resolve out the emission from
the large-scale molecular cloud. However, they may also resolve out
emission from the Class 0 envelopes.  This is caused by limited
{\it uv} coverage, particularly due to the absence of short baselines
and zero-spacing.  For these reasons, interferometers with good {\it
uv} coverage are required.  The recently commissioned
Combined Array for Research in Millimeter-wave Astronomy (CARMA)
provides the best opportunity with its unprecedented compact
configuration and image fidelity \citep{woody2004}.

In this paper, we present dust opacity spectral indexes $\beta$ of
Class 0 sources (L1448 IRS 2, L1448 IRS 3, and L1157) in
order to tackle when the dust grain growth responsible for $\beta \sim 1$ mainly occurs: before
or after the Class 0 stage. We do a parametric modeling in {\it uv}
space to address the $\beta$ values, as well
as image and visibility comparisons.  In addition, we examine
power-law density indexes via modeling.  First, we discuss 
our observations and data reduction, focusing on how well our CARMA
data incorporate with this study.  Afterward, we show our results in
images, {\it uv} visibilities, and visibility modelings.
At the end, we discuss the implications of our results.

\section{Target YSOs}

We have carried out observations of three Class 0 YSO regions (L1448
IRS 2, L1448 IRS 3, and L1157) using CARMA in the \omm\ and 2.7 mm
continuum. These three targets are well defined as Class 0 YSOs by
previous studies \citep[e.g.,][]{shirley2000,olinger1999}.  L1448
IRS 2 and IRS 3 are located in the dark cloud L1448 of the Perseus
molecular cloud complex at a distance of 250 pc.  They were first
revealed by IRAS observations \citep{bachiller1986}.  L1448 IRS 3
is the brightest infrared source in the dark cloud and has three
Class 0 sources (3A, 3B, and 3C), revealed by radio interferometric
observations \citep[]{curiel1990,terebey1997,looney2000}.
\citet{kwon2006} also studied the binary system of 3A and 3B, the
two interacting bipolar outflows, and the magnetic field in the
region, using polarimetric observations of the Berkeley Illinois
Maryland Association (BIMA) array in the \omm\ continuum and CO
\jto\ transition line.

On the other hand, L1448 IRS 2 at $\sim 3 \arcmin$ west of IRS 3
has not been focused on very much due to its weaker brightness.
However, \citet{olinger1999} identified it as a Class 0 YSO, using
far-infrared up to millimeter continuum observations.  In addition,
recent deep Spitzer Space Telescope (SST) IRAC observations have
shown a large bipolar outflow spanning over $5 \arcmin$ \citep{tobin2007}.
CARMA observations in CO \jto\ and \joz\ transitions also show a
well-developed bipolar outflow \citep{kwon_preparation}.

L1157 is a dark cloud in Cepheus.  The distance is not well known
but it is arguably about 250 pc \citep{looney2007}. Its envelope and
large bipolar outflow have been studied by radio single dish and
interferometric observations
\citep[e.g.,][]{bachiller2001,gueth2003,beltran2004}. The bipolar
outflow is known as chemically active, since various molecules have
been detected and interestingly there is an abundance gradient that
cannot be explained purely by excitation temperature differences
\citep{bachiller2001}. Recently, a flattened envelope has been
detected in absorption against polycyclic aromatic hydrocarbon
(PAH) background emission by deep SST IRAC observations \citep{looney2007}.

\section{Observations and Data Reduction}
\label{sec_obs}

We have carried out $\lambda = $ 1.3 mm and 2.7 mm continuum
observations towards three Class 0 sources, L1448 IRS 2, L1448 IRS
3, and L1157, using CARMA \citep{woody2004}, which is a recently
commissioned millimeter array, combining the BIMA and OVRO
(Owens Valley Radio Observatory).  It consists of 6 elements of 10.4
m antennas and 9 elements of 6.1 m antennas.\footnote{Recently 8 elements
of 3.5 m antennas (the Sunyaev-Zel$'$dovich Array) have been merged as well.} 
In order to achieve
a similar synthesized beam at the two wavelengths, the
$\lambda = $ 1.3 mm and 2.7 mm continuum data have been taken in
the most compact E configuration and the D configuration, respectively.
These two combinations of wavelengths and array configurations
provide well matched synthesized beams, about $5 \arcsec \times 5 \arcsec$.

This moderately matched beam size at these two wavelengths has 
not been achievable before
CARMA.  In interferometric observations, while high angular resolution
can be obtained via increasing baselines of antenna elements, there
is the usual missing flux problem.  This is because interferometric
observations are only sensitive to size scales corresponding to the
{\it uv} coverage.  To mitigate the missing flux issue, we need
either an additive single dish observation or well-defined {\it uv}
coverage with short baselines.  From this point of view, the most
compact CARMA E configuration is just right to study Class 0 envelope
structures, since the canonical size of Class 0 source envelopes
is several thousands of AU corresponding to a few tens of arc-seconds
in most nearby star forming regions (e.g., the Perseus molecular
cloud at a distance of 250 pc).  The E configuration provides
baselines from $\sim 6$ m to $\sim 60$ m ($\sim 4.6-46$ \klambda\
at \omm), which result in a synthesized beam (angular resolution)
of about $5 \arcsec \times 5 \arcsec$.  A simulation shows that our
data {\it uv} coverage recovers fluxes well ($> 50$\%) towards
extended features about up to 4 times the synthesized beam size.

CARMA has a couple of special features to realize the most compact
E configuration.  One is an anti-collision system installed on the
6.1 m antennas, which are located in the inner region of the
configuration.  Antennas stop whenever they are in a danger of
collision.  The other feature is the coordinated movement. In larger
configurations, D, C, and B configurations, antennas diagonally
move (simultaneously in azimuth and elevation) to reach a target.
However, in E configuration they go to a high elevation first and
move in azimuth followed by a movement to arrive at a designated
elevation, to reduce the collisional situations.

The $\lambda = 2.7$ mm continuum was observed in the D-like
commissioning configuration of 2006 fall and winter and D configuration
of 2007 summer, while the $\lambda = 1.3$ mm continuum was obtained
in the E configuration of 2007 summer.  Each data set was taken
with one or two double-side bands of a 500 MHz bandwidth in each
single-side band for the continuum observations.  Two or one extra
bands  were assigned to a CO rotational transition (\jto\ or \joz).
The CO rotational transition data are presented in another paper
with other molecular transition data.  The details of each observation
are listed in Table \ref{tab_obs}.  Two and three pointing mosaics
have been done to better cover the larger bipolar outflow regions
for the \cojto\ transition towards L1448 IRS 3 and L1157, respectively,
at $\lambda = 1.3$ mm.  For this study, the northwest pointing data
of L1448 IRS 3 and the central pointing data of L1157 were used.

The Multichannel Image Reconstruction, Image Analysis, and Display
\citep[MIRIAD,][]{sault1995} tools have been employed to reduce and
analyze data.  In addition to normal procedures (linelength, bandpass,
flux, and gain calibrations), shadow-defected data have been flagged
in the E configuration data.  Shadowing indicates cases of an
antenna's line-of-sight interrupted by other antennas and usually appears
in low elevation observations of compact configurations. The normal
effects of shadowing are reduction and distortion of incident
antenna power and abnormal gain jumps. Therefore, to obtain
reliable results the shadow-defected data were flagged in
the compact E configuration.

Further special attention needs to be given on flux calibration
for studies involving flux comparison between different
wavelengths like this study. To minimize errors caused by
primary flux calibrators, we used the same flux calibrator (Uranus)
at both wavelengths except L1157, which used MWC349 at
\omm\ and Mars at \tmm.  We expect 15\% and 10\% uncertainties
of flux calibrations at \omm\ and 2.7 mm, respectively, based on
the CARMA commissioning task of flux calibration.  During a
commissioning period extending to longer than 4 months, 12 calibrator
(quasar) fluxes had been monitored by CARMA.  As a result, the least
varying case showed about 13\% deviation in flux.  When considering the
intrinsic variability of quasars, it is expected that CARMA flux
calibrations have about $10-15$\% uncertainties.  As a result, we 
consider 15\% and 10\% uncertainties at \omm\ and 2.7 mm, respectively.  

In addition, we make synthesized beam sizes the same as possible at
both wavelengths, using weighting and tapering schemes, in order
to minimize the beam size effect on the flux comparison.  
After proper weighting and tapering schemes, we could
match the beam sizes to within 1\%.  The details of applied weighting
and tapering schemes are listed in Table \ref{tab_beam} with final
synthesized beams.  Briggs' robust parameter
is used \citep{briggs1995}, which is a knob to provide intermediate
weighting between natural and uniform weighting.  The parameter
of $2$ gives a weighting close to natural weighting and $-2$
close to uniform weighting.

\vspace{0.3cm}
\section{Observation Results}

\subsection{Dust opacity spectral index maps}
\label{sec_betamap}

Total flux ($F_\nu$) of the thermal dust continuum emission represents
the total mass ($M_T$) of the source, if the source is optically thin 
at the observational frequencies,
\begin{equation}
F_\nu \approx \kappa_\nu\, B_\nu(T_{d})\, \frac{M_T}{D^2},
\end{equation}
where $\kappa_\nu$, $B_\nu(T_{d})$, $M_T$, and $D$ are opacity 
(mass absorption coefficient)
of the dust grains, blackbody radiation intensity of a dust temperature
$T_{d}$, total mass, and distance to the source, respectively.
The opacity of dust grains ($\kappa_{\nu}$) depends on dust
properties such as sizes, components, and shapes.  If the dependence
is simple, for example a power law ($\kappa_\nu \varpropto \nu^\beta$),
the dust grain properties can be studied by observations at two
frequencies.  In addition,  in the case that the Rayleigh-Jeans approximation
of blackbody radiation is applicable ($h\nu \ll kT$), the relationship
between spectral indexes of the observed flux densities ($\alpha$) and 
spectral indexes of the dust grain opacity ($\beta$) is simply,
\begin{eqnarray}
F_\nu &\approx& F_{\nu_0} \Big(\frac{\nu}{\nu_0}\Big)^\alpha \nonumber \\
F_\nu &\approx& \kappa_{\nu}\, B_\nu(T_d)\, \frac{M_T}{D^2} \nonumber \\
      &\approx& \kappa_{\nu_0}\Big(\frac{\nu}{\nu_0}\Big)^\beta\,\, 
                \frac{2kT_d}{c^2}\, \nu^2\, \frac{M_T}{D^2} \nonumber \\
\textrm{therefore}~~~~~ \alpha &\approx& \beta + 2.
\end{eqnarray}
Note that this relation is valid only in the optically thin assumption and
the Rayleigh-Jeans approximation.

\citet{draine2006} showed that $\beta$ mainly
depends on the size distribution of dust grains rather than their
components and shapes; small $\beta$
($\sim 1$) is likely indicating dust grain size distribution up to
$3 \lambda$. Since our observations are up to 3 mm, $\beta \sim 1$
would suggest a grain size distribution up to about 1 cm.

\begin{figure}
\begin{center}
\includegraphics[angle=270,scale=0.85]{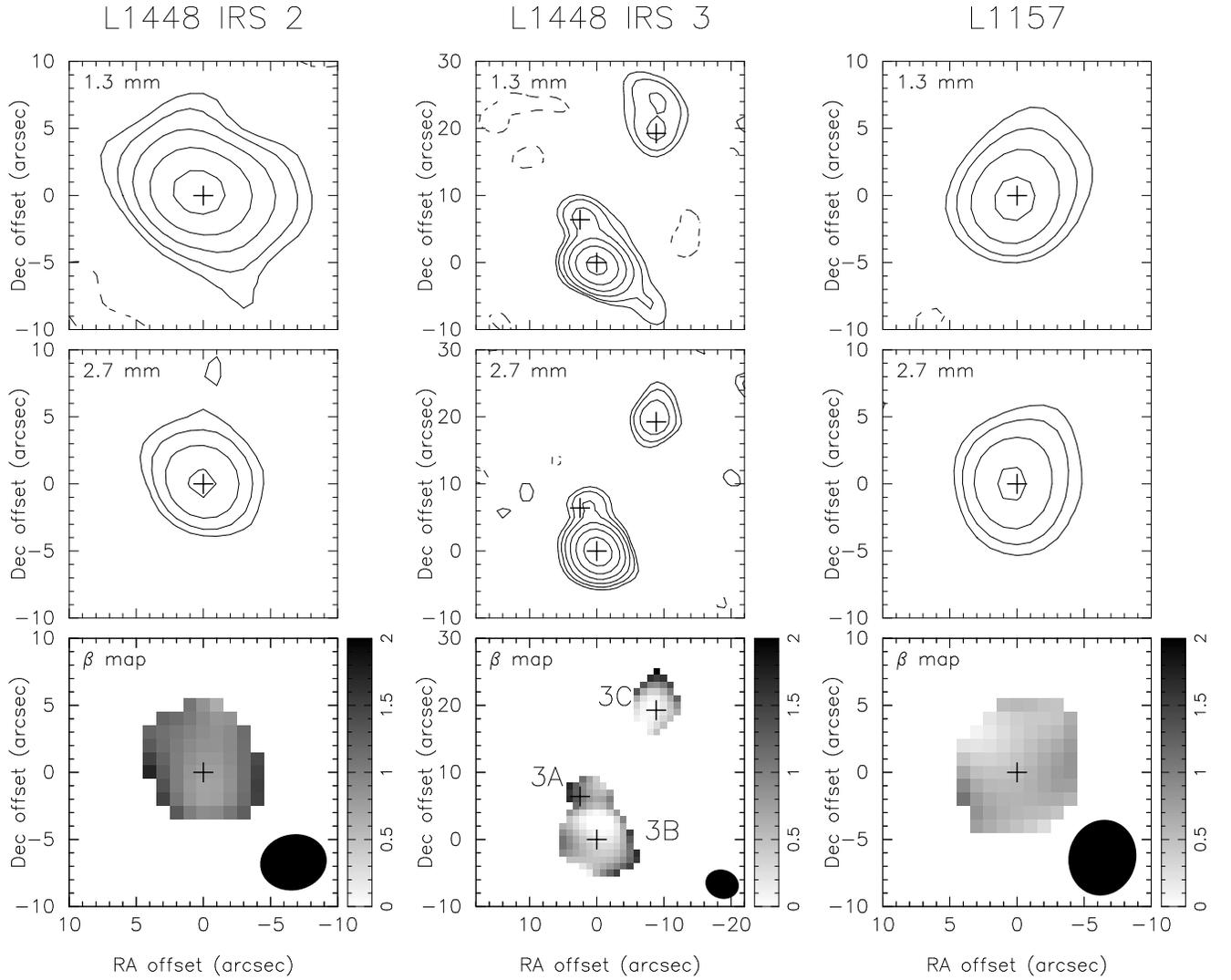}
\caption{Dust continuum in \omm\ and $2.7$ mm and dust opacity 
spectral index ($\beta$) maps of L1448 IRS 2, L1448 IRS 3, and L1157. 
Note that most $\beta$ values are less
than 1. The gray scales are the same in all three maps, although the
values are distributed in different ranges.  The statistics of the
values are in Tab. \ref{tab_beta}.  Synthesized beams are the same in all
three maps towards each target (Tab. \ref{tab_beam}) and 
plotted on the bottom right of the $\beta$ maps.
The L1448 IRS 3A, 3B, and 3C positions came from \citet{looney2000}
and the \omm\ map of L1448 IRS 3 was re-centered; the pointing center
was $\textrm{(RA, Dec)} \approx (-4 \arcsec, 8 \arcsec)$.
The contours of dust continuum maps are 3, 5, 9, 17, 33, and 65
times $\sigma = 3.4$ and 1.1 mJy beam$^{-1}$ (\omm\ and 2.7 mm maps
of L1448 IRS 2), 10 and 1.6 mJy beam$^{-1}$ (\omm\ and 2.7 mm maps
of L1448 IRS 3), and 13 and 2.4 mJy beam$^{-1}$ (\omm\ and 2.7 mm
maps of L1157).  
\label{fig_betamap}} 
\end{center} 
\end{figure}

Figure \ref{fig_betamap} presents maps of L1448 IRS
2, L1448 IRS 3, and L1157.  Dust continuum maps at \omm\ and
\tmm\ have been separately constructed using different weightings
and taperings as described in \S\ \ref{sec_obs} and Table \ref{tab_beam}
in order to have as similar synthesized beams as possible at the
two wavelengths.  Afterwards $\beta$ values of each source have been
calculated using the two continuum images.  Only regions above three
signal-to-noise ratio (SNR) levels on the both maps have been used to derive
$\beta$ assuming
\begin{equation}
\label{eq_beta}
\beta = \frac{\log(F(\nu_1)/F(\nu_0))}{\log(\nu_1/\nu_0)} - 2,
\end{equation}
where $\nu_1$ and $\nu_0$ are frequencies corresponding to \omm\
and \tmm\ data, as listed in Table \ref{tab_beam}.  Note
that the Rayleigh-Jeans approximation and the optically thin assumption
are used.  In the case of an average dust temperature of
about 30 K, the upper limit of frequencies to which the Rayleigh-Jeans
approximation can be applied is about 625 GHz. Since the higher
frequency of our data is about 230 GHz, the assumption is valid for
this study.  However, caution should be taken in $\beta$ comparison
at submillimeter wavelengths for cold objects such as the Class 0
YSO envelopes.

As shown in Figure \ref{fig_betamap}, most $\beta$ values in the
three targets are less than 1.  For a convenient comparison, the
same gray scales have been adopted for all three maps. The
actual ranges of $\beta$ values are in Table \ref{tab_beta} with
the averages. As listed in the table, the maximum values are larger
than 1.0.  However, those large $\beta$ values appear only on a few
pixels of source boundaries, which 
may be due to contamination from ambient clouds.  
$\beta$ and its averages in most regions of the three sources are
similar to or less than 1.  In the case of L1448 IRS 3, in which
three Class 0 sources (3A, 3B, and 3C) exist, $\beta$ values
corresponding to the three sources are separately listed in Table
\ref{tab_beta}.  Like the other targets, these three sources
of L1448 IRS 3 have $\beta$ around or less than 1. The L1448 IRS
3A and 3B fluxes are obtained simply by cutting the protuberance in
Figure \ref{fig_betamap}.
Table \ref{tab_beta} also has $\beta$ values obtained from the total
fluxes at the two wavelengths, which have been estimated in source
regions limited by the three SNR threshold at both
wavelengths. All sources except L1448 IRS 3B have $\beta$ values
comparable to the mean values of the $\beta$ maps.  

Another feature to note is that there are $\beta$ gradients with
radius in all sources.  L1157 has a smaller $\beta$ in the
northeast-to-southwest direction, roughly consistent with the \omm\
and 2.7 mm results of \citet{beltran2004}.  However, it is noteworthy
that they restored their two images with an identical beam size without
any weighting schemes, which could cause a biased result due to
different {\it uv} coverage of the two wavelength data.  The radial
dependence of $\beta$ is better shown in \S\ \ref{sec_vis_comp} and is
discussed in detail for the L1448 IRS 3B case via modeling in \S\
\ref{sec_discussion}

\subsection{Visibility data comparison}
\label{sec_vis_comp}

We have also examined $\beta$ values in {\it uv} space, which is
the Fourier transformed space of an image.  Data of
interferometric observations are obtained in {\it uv} space and
called {\it uv} visibilities or just visibilities.  To obtain a sky
intensity distribution, inverse Fourier transformation and deconvolution
(e.g., CLEANING algorithm) are employed \citep[e.g.,][]{isra2001}.  However,
limited {\it uv} coverage causes difficulties, i.e., the deconvolution
introduces systematic biases, especially for non-point, extended sources.  
One of the best ways to overcome this difficulty is to
investigate the visibility data in {\it uv} space instead.

\begin{figure}
\includegraphics[angle=270,scale=0.70]{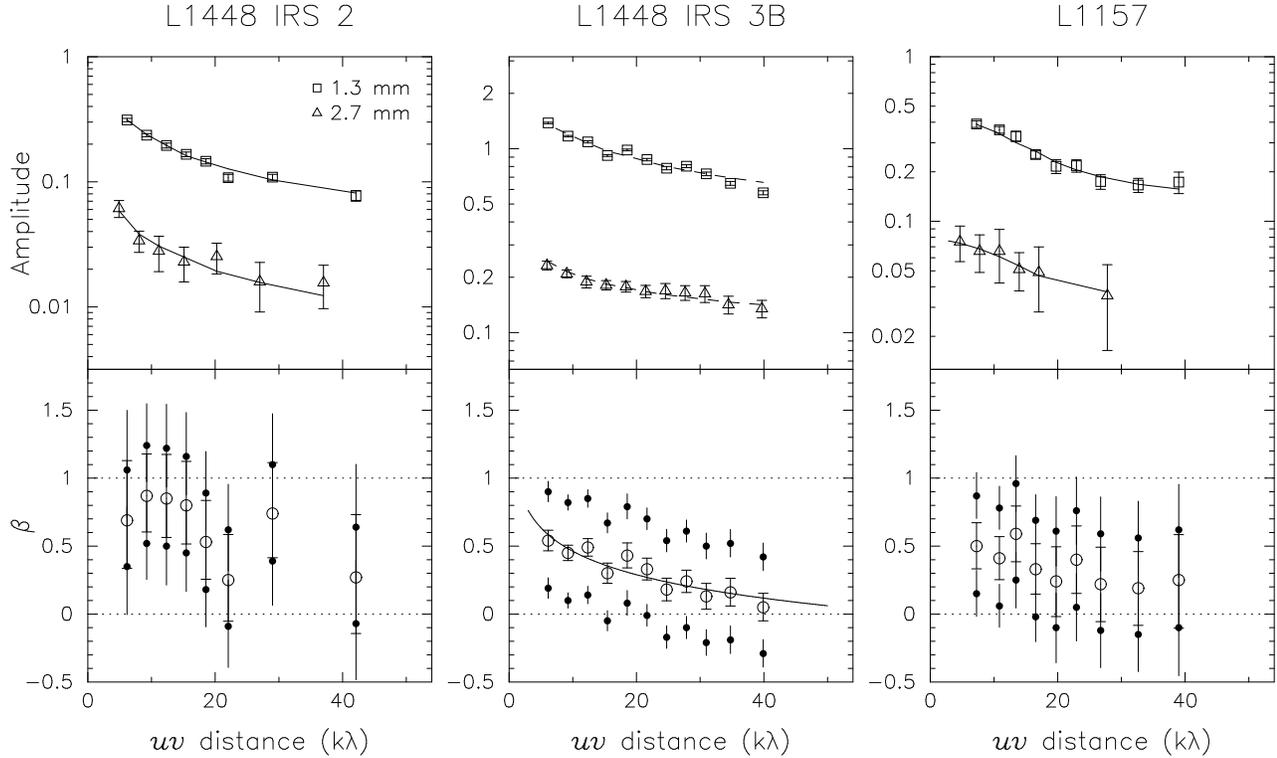}
\caption{Amplitude (upper panels) and dust opacity spectral index 
$\beta$ plots (lower panels) of the
three targets, L1448 IRS 2, L1448 IRS 3B, and L1157, along 
{\it uv} distance. The open squares present $\lambda = 1.3$ mm
data and the open triangles are for $\lambda = 2.7$ mm data.
The error bars in the amplitude plots are
statistical standard errors of visibilities in each bin.
The solid and dashed lines present the best fit models described
in \S\ \ref{sec_modeling} and Fig. \ref{fig_likelihood}.
The open circles and error bars with caps in the $\beta$ plots 
indicate $\beta$ values and distribution regions 
corresponding to the amplitude statistical errors.
The filled circles and error bars without caps present cases
assuming 15\% higher amplitudes at $\lambda = 1.3$ mm and 10\% 
lower amplitudes at $\lambda = 2.7$ mm (resulting in the largest $\beta$
case within absolute flux calibration uncertainty) and 15\% lower
at $\lambda = 1.3$ mm and 10\% higher at $\lambda = 2.7$ mm 
(resulting in the smallest $\beta$ case). 
The $\beta$ values are calculated at the {\it uv} distance bin centers of the \omm\
data.  The visibilities of \tmm\ at the positions are interpolated
linearly using nearest bin values and in the case of extrapolation
the nearest bin values are assumed.  The solid line in the $\beta$ plot
of L1448 IRS 3B is a logarithmic fit to the data.  
Refer to the text for further details.
\label{fig_uvamp}} 
\end{figure}

The results of $\beta$ calculated in {\it uv} space are displayed
in Figure \ref{fig_uvamp}.  Visibilities have been vector-averaged
in annuli.  Since the envelope structures from our observations are
spherical, the annulus averaging is valid.  The annulus bin sizes
are $\sim 3.1$ \klambda\ except when the SNR is
too low, usually at the relatively longer baselines.
This is most noticeable in L1157 at \tmm.
Although the {\it uv} coverage is comparable at both
wavelengths, the lower SNR at \tmm\ requires larger bins.
The $\beta$ values are calculated at the \omm\ bins with \tmm\ visibilities
linearly interpolated using the nearest bin values.  When the \omm\ bin center
is beyond last \tmm\ bin center (extrapolation case), then the nearest bin value for \tmm\
is used.

In the case of L1448 IRS 3, only 3B is
considered for the $\beta$ calculation in {\it uv} space.  The other
two objects, 3A and 3C, are too small and weak to carry out the
calculation.  On the other hand, 3A and 3C should be removed from the
visibilities to obtain the 3B data.  Using the MIRIAD
task UVMODEL and image models excluding the two components, we
subtracted the 3A and 3C visibilities at both \omm\ and \tmm\
separately.  In addition, since the \omm\ data set has been taken
with two pointings offset from the center, we compensated the primary
beam sensitivity loss using a UVMODEL multiplication.

In Figure \ref{fig_uvamp}, the upper panels show amplitudes of \omm\
(open squares) and \tmm\ cases (open triangles).
The error bars represent the statistical standard errors in each bin.  
The solid and dashed lines present the best fit models described
in \S\ \ref{sec_modeling} and Figure \ref{fig_likelihood}.
The lower panels show $\beta$ values with {\it uv} distance,
calculated by equation (\ref{eq_beta}). The open circles indicate
$\beta$ values calculated from the {\it uv} visibilities shown on
the upper panels. The error bars with caps on the open circles
represent $\beta$ value ranges corresponding to the statistical
amplitude errors of the upper panels. The filled circles and error
bars without caps show the effect that the absolute flux calibration 
uncertainty has on the calculation of $\beta$.
We adopt 15\% flux calibration uncertainties for \omm\ data and
10\% for \tmm\ data, as discussed in \S\ \ref{sec_obs}.  The larger
$\beta$ points indicate the case in which 15\% higher fluxes at
\omm\ and 10\% lower fluxes at \tmm\ are considered and vise verse
for the lower $\beta$ points.  The $\beta$ ranges are around $\pm
0.35$, as $\textrm{log}(1.15/0.90)/\textrm{log}(\nu_1/\nu_0) \approx
0.35$ where $\nu_1/\nu_0 \approx 2$ (refer to eq. \ref{eq_beta}).

Two main features should be noted in Figure \ref{fig_uvamp}.  One
point is that the $\beta$ values are around 1 or less than 1 in all
three objects. It is arguably true even when considering the absolute
flux calibration uncertainties.  The other point is the radial dependences
of $\beta$.  In L1448 IRS 2 and L1157, $\beta$ arguably decreases
on smaller scales (larger {\it uv} distances).  L1448 IRS 3B,
however, distinctly presents a radial dependence.  
The $\beta$ variation is fit with 
the logarithmic function of 
$\beta(\zeta) = 1.0 - 0.57~\textrm{log}(\zeta)$,
where $\zeta$ is the {\it uv} distance in units of k$\lambda$.
When assuming power-law distributions of density and temperature of
envelopes as discussed in \S\ \ref{sec_modeling}, 
the distributions of the intensity integrated 
along line-of-sight as well as the radial intensity follow a power-law
under the optically thin assumption and Rayleigh-Jeans approximation 
\citep{adams1991}.  When ignoring primary beam effects of interferometers
and assuming infinite size envelopes, the visibilities are also in
a power-law \citep[e.g.][]{harvey2003,looney2003}.  
As $\beta$ is obtained from equation (\ref{eq_beta}) here,
we assume a logarithmic function of $\beta(\zeta)$.
There are a few possible interpretations to explain this radial
dependence of $\beta$.  It could be caused by increasing the fraction
of optically thick emission on smaller scales due to the denser
central region. \citet{beckwith1990} discussed that the optically
thick emission fraction ($\Delta$) decreases $\beta$ by a factor
of ($1+\Delta$), i.e., $\beta \simeq \beta_0 / (1+\Delta)$.  Similarly,
it could be due to an optically thick, unresolved, deeply embedded
disk structure at the center.  On the other hand, it could indicate
a faster grain growth in the denser central region or dust grain
segregation suggested by some star formation theories, for example,
ambipolar diffusion in magnetically supported molecular cloud
\citep{ciolek1996}.  The radial dependence is discussed in more
detail in \S\ \ref{sec_discussion}.

\section{Modeling in {\it uv} space}
\label{sec_modeling}

As mentioned in \S\ \ref{sec_vis_comp}, images of extended features
constructed from interferometric observations may be biased due to
limited {\it uv} coverage. 
In contrast, comparing visibility data against source models
transformed to the visibility plane (including 
the primary beam modification, Fourier transformation, and visibility sampling),
is not prone to these imaging deconvolution biases.
Therefore, we carry out envelope
modeling in {\it uv} space rather than in image space. 
In other words, we compare observation visibilities with model 
visibilities sampled over the observation {\it uv} coverage,
after obtaining {\it uv} models by the Fourier-transformation
of image models. 

We assume that the temperature distribution of dust grains is 
in radiative equilibrium with the central protostar, ignoring 
heating by gas and cosmic rays \citep[][p 193]{spitzer1978ppim}:
\begin{equation}
\label{eq_radeq}
c \int_0^\infty Q_a(\nu)\, u_\nu\, d\nu = 4\pi \int_0^\infty 
Q_a(\nu)\, B_\nu(T_d)\, d\nu,
\end{equation}
where $Q_a(\nu)$, $u_\nu$, $B_\nu(T_d)$, and c are absorption
efficiency factor, radiation energy density, black body radiation
intensity of temperature $T_d$, and speed of light, respectively.
The radiation energy density ($u_\nu$) at a distance $r$ from the
center can be expressed as $\pi B_\nu(T_*)\, R_*^2/r^2$, where $T_*$
and $R_*$ are an effective temperature and a radius of a central
protostar. Assuming $Q_a(\nu) \propto \nu^\beta$, equation (\ref{eq_radeq})
gives a temperature distribution of dust grains \citep{beckwith1990}, 
\begin{equation}
T_d(r)= T_*\, \Big( \frac{1}{2}\frac{R_{\ast}}{r} \Big) ^{2/(4+\beta)}.
\end{equation}
Again, $\beta$ is the dust grain opacity spectral index 
($\kappa_{\nu} = \kappa_0 (\nu/\nu_0)^{\beta}$).
This equation can also be formulated with a grain temperature $T_0$ 
at a distance $R_0$ from the central protostellar luminosity $L_0$, 
as \citep[e.g.,][]{looney2003}
\begin{equation}
T_d(r) = T_0\, \Big( \frac{R_0}{r} \Big) ^{2/(4+\beta)} 
\Big( \frac{L_*}{L_0} \Big) ^{1/(4+\beta)}. 
\end{equation}
Although the inner region, which might be optically
thick, could have a sharper temperature gradient than this relation
\citep[e.g.,][]{wolfire1986,looney2003}, it is limited at the very
central region, and our results are not sensitive to the possibility,
as further discussed in \S\ \ref{sec_discussion}.

Some previous studies \citep[e.g.,][]{harvey2003} considered the
external heating by the interstellar radiation field, using a
temperature lower limit of 10 K.  However, we do not explicitly
include this effect, since the temperature lower limit is uncertain
and the lowest temperature of our modeling is comparable, about 7.3
K at $r=7000$ AU when adopting $T_0=100$ K at $r=10$ AU.  In addition,
tests show that the temperature lower limit does not
change our results, as previous studies also reported
\citep[e.g.,][]{harvey2003}.  The outer envelope heated externally
by the interstellar radiation field would be the main intensity
component in sources without any central heating objects, but in
Class 0 YSOs the central high
temperature region drives the emission.  Besides, interferometric
observations are not so sensitive to the outer envelope,
where the effect of the temperature lower limit is largest.

The power-law density distribution is assumed for envelopes, $\rho(r) = 
\rho_0 (r/r_0)^{-p}$.
Therefore, the intensity of envelopes on the plane of the sky is 
calculated as
\begin{equation}
\label{eq_I}
I_{\nu} = \int B_{\nu}(T_d(r))~e^{-\tau_\nu}~d\tau_\nu
= \int B_{\nu}(T_d(r))~e^{-\tau_\nu}~\rho(r)~\kappa_\nu~dL,
\end{equation}
where L indicates the line-of-sight from the observer and the optical depth 
$\tau_{\nu} = \int_0^L \kappa_{\nu}~\rho(r)~dL'$. Spherical envelopes
with an outer radius of $R_{out}$ and with an inner hole of a radius
of $R_{in}$ are assumed. Therefore, the density distribution can
be expressed with the total envelope mass $M_T$ (when $p \neq 3$) as
\begin{eqnarray}
M_T &=& \int_{R_{in}}^{R_{out}} \rho(r)~4\pi r^2~dr \nonumber \\
 &=& \frac{4\pi}{3-p} (R_{out}\,\!^{3-p}-R_{in}\,\!^{3-p})~\rho_0\,r_0\,\!^p \\
\rho(r) &=& \rho_0\,r_0\,\!^p~r^{-p} \nonumber \\
 &=& M_T~\frac{3-p}{4\pi}(R_{out}\,\!^{3-p}-R_{in}\,\!^{3-p})^{-1}~r^{-p}.
\end{eqnarray}
Substituting the density expression with the total envelope mass
into the optical depth of equation (\ref{eq_I}) shows a coupling
of $M_T$ and $\kappa_0$ --- in the case that the envelope is optically 
thin and the Rayleigh-Jeans approximation ($B_\nu(T_d(r)) \approx
2kT_d(r)/\lambda^2$) is valid, $T_0$ is also coupled.  Normally the
envelopes of this stage YSO are optically thin in the \omm\ and
$2.7$ mm continua except the very central regions (within a few
tens of AU) and the Rayleigh-Jeans approximation is applicable,
which means that the $M_T$, $\kappa_0$, and $T_0$ are all likely
coupled.  However, note that the optically thin assumption and
Rayleigh-Jeans approximation, which are assumed in $\beta$ calculations
of observational data in \S\ \ref{sec_betamap} and \S\ \ref{sec_vis_comp},
are not assumed in the modeling to avoid biases. Here we just
intend to point out that the three parameters are likely to be
coupled.

After constructing intensity
image models, they are corrected by three different CARMA primary
beams, which correspond to baselines of two 10.4 m antennas,
two 6.1 m antennas, and 10.4 m and 6.1 m antennas. The three primary-beam
corrected images are Fourier-transformed into {\it
uv} space and model visibilities are sampled over the actual
observational {\it uv} coverage of the three different baselines.
Comparison between model and observation visibilities is done by
vector averaged values in annulus bins.  Although bipolar outflows 
at this stage carve a cavity \citep[e.g.,][]{seale2008}, 
the effect is minor \citep{chandler2000}, 
especially at our intermediate angular resolution.  
In addition to the bipolar outflow effect, envelopes might be clumpy.
However, the effect on our modeling is also insignificant, since
the angular resolution of our data is intermediate and annulus-averaged
values are used for the comparison of models and data.

Parameters involved in our modeling are $p$ (power-law density
index), $\beta$ (opacity spectral index), $M_T$ (envelope total
mass), $\kappa_0$ (opacity coefficient at $\nu_0$), $T_0$ 
(grain temperature at $R_0$), $R_{in}$ and 
$R_{out}$ (inner and outer radii of envelopes), and $F_{pt}$ (a
central point source flux at \tmm).  Among these, two parameters
are fixed: $\kappa_0 = 0.0114$ cm$^2$ g$^{-1}$ at $\nu_0 = 230$ GHz
and $T_0 = 100$ K at $R_0 = 10$ AU.  As discussed, the $\kappa_0$ 
and $M_T$ are coupled (and so $T_0$ is mostly), so we cannot well
constrain these parameters simultaneously.  The $T_0$ at $R_0$ corresponds
to a central object luminosity of 1.67 $L_\sun$ and the $\kappa_0$
at $\nu=230~\textrm{GHz}$ is the average of $\beta=1$ and $2$ cases
in $\kappa_\nu = 0.1\,(\nu/1200~\textrm{GHz})^\beta$, assuming
a gas-to-dust mass ratio of 100
\citep[e.g.,][]{hildebrand1983,beckwith1990}.  \citet{ossenkopf1994}
also reported $\kappa \approx 0.01$ cm$^2$ g$^{-1}$ at \omm\ for
dense protostellar cores via dust coagulation model calculation,
when using a gas-to-dust mass ratio of 100.
Note that $\kappa_0$ is not very well known and has a large uncertainty
\citep[e.g.,][]{hildebrand1983,beckwith1991} so we need to pay
attention to the fact that the total mass $M_T$ could have a large
uncertainty.  $M_T$ can also be scaled by the presumed $T_0$.  

The central point source flux ($F_{pt}$) is designed to simulate
an unresolved central disk structure. We assumed that the point
sources are optically thick so that the flux density spectral index
is 2 under the Rayleigh-Jeans approximation, meaning $\beta = 0$.
In the case of L1448 IRS 2 there is no point source required, since
there is no flat visibility amplitude on the small scales, particularly
at \omm\ in Figure \ref{fig_uvamp}. It may indicate that the central
disk structure of the source is not so significant.  In contrast,
L1157 has a flat profile on the small scales, which means a compact
structure at the center.  Therefore, a point source is adopted to
fit the data.  Indeed, \citet{beltran2004} reported a compact
component (size $< 1 \arcsec$) of 25 mJy and 78 mJy at \tmm\ and 1.3 mm,
respectively.  On the other hand, the point source of L1448 IRS 3B
was applied for a different reason: to simulate a radial dependence
of $\beta$.  As shown in Figure \ref{fig_uvamp}, there is a clear
radial dependence of $\beta$, which results in no good fits with a
constant $\beta$ over all scales.  It is why an optically thick
point source is considered, although there is no flat feature on
the small scales.  Note that even higher angular resolution
observations have not detected such a point source signature
\citep{looney2003}.  We further discuss the $\beta$ radial dependence of
L1448 IRS 3B in \S\ \ref{sec_discussion}.

In order to find good fit models, we search grids of parameters,
$p$, $\beta$, $M_T$, $R_{in}$, $R_{out}$, and $F_{pt}$.  Parameter
set information of the three sources is listed in Table \ref{tab_param}.
On each grid point of parameters, the reduced $\chi^2$ ($\chi_\nu^2$)
has been calculated. The two wavelength data were used simultaneously
for fitting.  Note that the absolute $\chi_\nu^2$ values
particularly in L1448 IRS 3B ($\sim 8.7$) are large, compared to
L1448 IRS 2 ($\sim 1.6$) and L1157 ($\sim 1.5$).  This is because the
relatively small standard errors due to the high brightness of L1448
IRS 3B make fitting very difficult. The L1448 IRS 3B data may have
imperfect exclusion of the companion L1448 IRS 3A, which
might cause a difficulty in fitting.  However, it is unlikely to be
the main effect, since the companion is relatively weak and we
subtracted the component as mentioned in \S\ \ref{sec_vis_comp}.  In
addition, the vector averaging in annuli minimizes the effect.  On
the other hand, it may indicate that the simple power-law model is
not appropriate to explain high SNR observations
\citep[e.g.,][]{chiang2008}.  

\begin{figure}
\includegraphics[angle=270,scale=0.70]{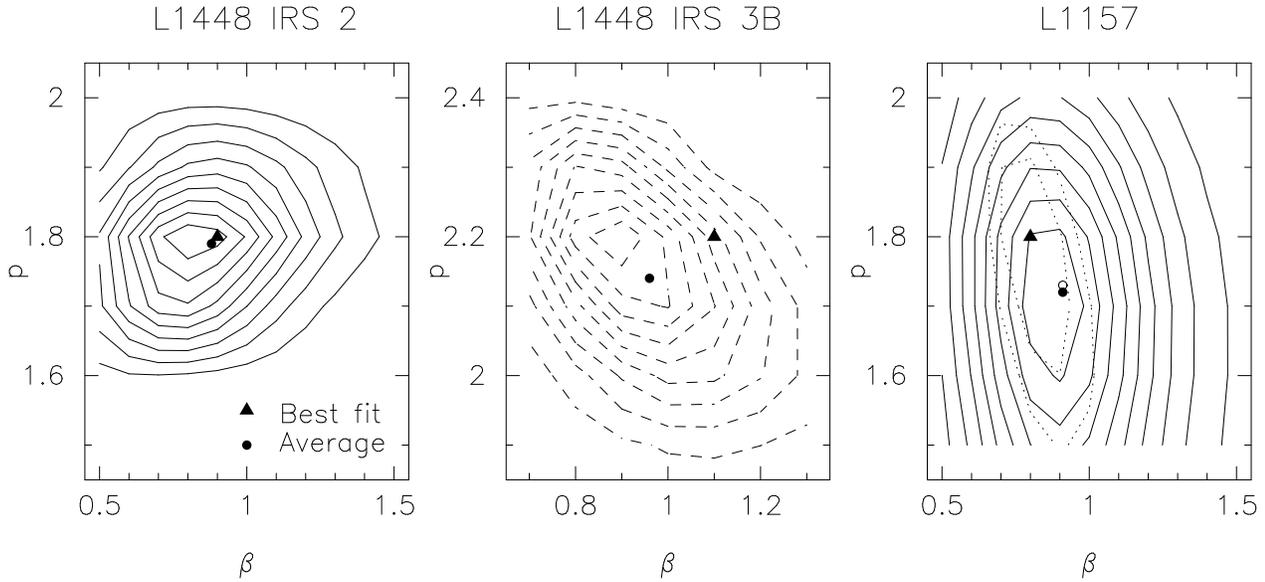}
\caption{Model fitting results of three Class 0 sources in likelihood. 
The contour levels are from 90\% of the peak value with steps of 10\%. 
The triangles mark best fit $p$ and $\beta$ pairs and
circles indicate likelihood weighted averages of $p$ and $\beta$.
To indicate that the model of L1448 IRS 3B is not the most reliable one in
this paper (refer to \S\ \ref{sec_discussion}), its contours are presented
by dashed lines.
The two dotted contours of L1157 indicate 90\% and 80\% of the peak
likelihood based on all models in the parameter ranges of Tab. \ref{tab_param}
and the solid contours are the likelihood distribution obtained from
better limited models. Refer to the text for details. In the two cases,
the best fit model is identical and the likelihood weighted averages are
slightly different. 
\label{fig_likelihood}}
\end{figure}

We adopt a likelihood calculation to constrain $p$ and $\beta$,
instead of reporting large ranges of each parameter to fit the data.
Reporting good fit parameter ranges could bias the impression of
the results, since each parameter value in the range comes from
different combinations of the other parameters.  
The likelihood function we adopt is $\textrm{exp}(-\chi_\nu^2/2)$, 
since the annulus averaged visibilities
have a Gaussian distribution based on the central limit theorem.  
As we want to constrain $p$ and $\beta$, the likelihoods of all
grid points with common $p$ and $\beta$ are summed. The sum now
indicates the likelihood of a set of $p$ and $\beta$. Finally, it
is normalized by the total sum of the likelihoods in each plot of
Figure \ref{fig_likelihood}, which means that the plots are comparable
to probability density distributions of $p$ and $\beta$.  Note that
we do not consider the absolute flux calibration uncertainties for
fitting.  In other words, we use data points marked with open symbols
in Figure \ref{fig_uvamp}.  Note that while systematic changes of
absolute fluxes in the same direction at both \omm\ and $2.7$ mm
affect the total mass $M_T$, the opposite direction changes mainly
influence $\beta$.  We estimate that the maximum $\beta$ ranges
caused by the absolute flux calibration uncertainties are $\pm
0.35$, as mentioned in \S\ \ref{sec_vis_comp}.

\begin{figure}
\begin{center}
\includegraphics[angle=270,scale=0.60]{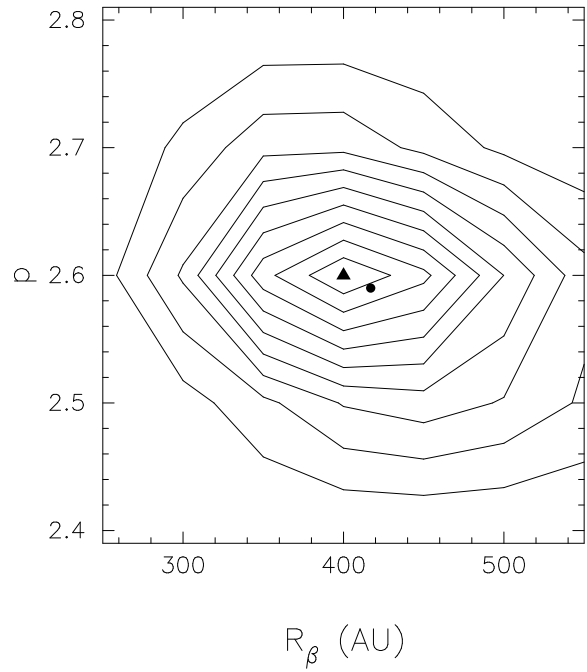}
\end{center}
\caption{Likelihood plot for models with variable $\beta$ along the 
envelope radius.  $R_\beta$ is the radius where $\beta = 1.7$ outward. 
Refer to the text for detailed discussions.
\label{fig_rbeta}}
\end{figure}

We present the most likely $\beta$ and $p$ in
Figure \ref{fig_likelihood}.  As clearly shown in the figure, $\beta$
of the three sources are most likely to be around 1 even in the
modeling without the optically thin assumption and Rayleigh-Jeans
approximation.  These are the first clear modeling results showing the
$\beta$ of Class 0 YSOs.  The contours in Figure \ref{fig_likelihood}
indicate likelihood levels of 90\% down to 10\% of the peak 
in steps of 10\% and the
triangles and circles mark the $p$ and $\beta$ pairs of the best
fit models and likelihood weighted averages of individual parameters, 
respectively.  Note that, therefore, the combinations of the weighted 
averages are not necessarily the best fit. Since
a model with a point source is not the best one for L1448 IRS 3B,
its contours are drawn in dashed lines.  (The best model is discussed
in \S\ \ref{sec_discussion} and displayed in Figure \ref{fig_rbeta}.)
The broader distribution in $p$ of L1157 is due to the adopted point
sources.  As having a point source implies a density gradient,
it lowers the density index.  The two dotted contours in the L1157
plot present 90\%
and 80\% of the peak likelihood based on all models in the whole
range of the point source fluxes $F_{pt}$ ($0.000 - 0.035$ Jy at
\tmm) listed in Table \ref{tab_param}.  In contrast, the solid
contours of L1157 in Figure \ref{fig_likelihood} show the likelihood
distribution obtained from models in a limited range of $F_{pt}$
($0.015-0.025$ Jy) around the likelihood weighted average ($0.019$
Jy at \tmm\ and $0.078$ Jy at \omm), which is consistent with the
compact component flux measured by \citet{beltran2004}.

While the power-law density indexes of L1448 IRS 2 and L1157 are
around 1.8 and 1.7, respectively, that of L1448 IRS 3B is around
2.1.  The density index of L1448 IRS 3B is consistent with the lower
limit of \citet{looney2003} using BIMA data and the L1157 result
is consistent with that of \citet{looney2007} using Spitzer IRAC
absorption features.  The density distribution of L1448 IRS 2 has
not been studied.  It is interesting to note that star formation
theories have suggested density indexes between 1.5 and 2.0;
``inside-out'' collapse models \citep[][]{shu1977} suggested 1.5
for the inside free-fall region and 2.0 for the outside isothermal
envelope, where the expansion wave does not reach yet, and ambipolar
diffusion models \citep[e.g.,][]{mouschovias1991,tassis2005b}
suggested around 1.7 but with the very inner regions dependent on
magnetically controlled accretion bursts.  Although we do not attempt
to constrain the star formation theories in this paper, the difference
in density indexes between L1448 IRS 3B and the others is noteworthy.
The difference even increases in the better model of L1448 IRS 3B in
\S\ \ref{sec_discussion}.

It is important to note that the constraints on the inner and outer
radii are not very strong.  While the inner radius of L1448 IRS 3B
is likely to be 10 AU rather than 20 AU, there is no likely inner
radius for L1448 IRS 2 and L1157 in the parameter search space.  In
addition, while the outer radius of L1157 is likely around $2000-2500$ AU,
the outer radii of L1448 IRS 2 and L1448 IRS 3B cannot be constrained
well due to lack of sensitivity of the data toward large scales.
We can only say that the preferred fits for these two sources
have a larger outer radius.
The values given in Table \ref{tab_param} are limited by
our parameter search space.

\section{Radial Dependence of $\beta$}
\label{sec_discussion}

We verify the radial dependence of $\beta$ that is shown in L1448
IRS 3B and attempt a modeling with $\beta$ as a function of radius
in this section.  This result is the first evidence to clearly show
a radial dependence of $\beta$ in Class 0 YSOs via {\it uv}
modeling.  Some previous studies have suggested a radial dependence
of $\beta$, for example, in dust cores of NGC 2024 \citep{visser1998},
the Class 0 source HH211-mm \citep{chandler2000}, and four Class I
sources \citep{hogerheijde2000}.  However, the results are not clear
and it could be due to other effects such as an optical thickness
effect or an improper consideration of temperature effects, since
their results are based on submillimeter wavelength observations,
in which the $\beta$ evaluation is more sensitive to the temperature.

\begin{figure}
\begin{center}
\includegraphics[angle=270,scale=0.60]{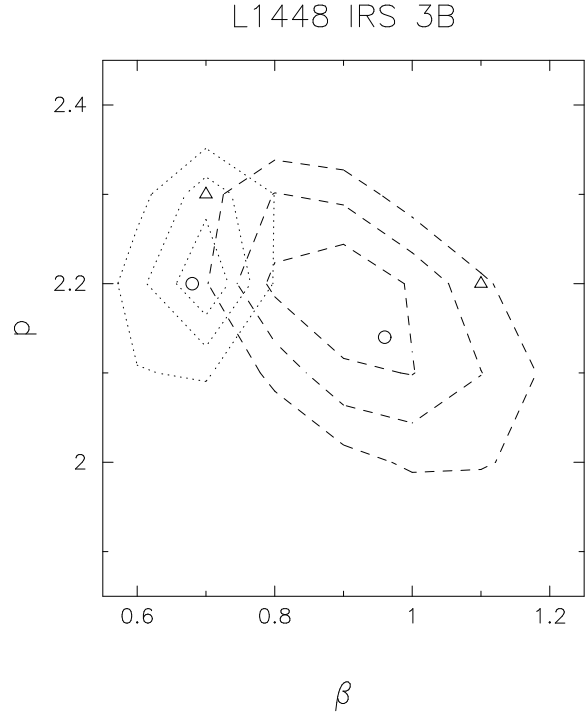}
\end{center}
\caption{Likelihood plots of two cases, (a) dashed contours: model
of a black body (optically thick) central point source same in 
Fig. \ref{fig_likelihood}
and (b) dotted contours: model of a central point source with a
$\beta$ same as the envelope.  Triangles presents best fit
values and circles indicate likelihood weighted average values. Note
that the best model of case (b) gives a worse fit ($\chi_\nu^2
\sim 11$) than case (a) ($\chi_\nu^2 \sim 8.7$).  The contour
levels are 80\%, 60\%, and 40\% of each likelihood peak value.
\label{fig_ptfluxbeta}} 
\end{figure}

As mentioned in \S\ \ref{sec_modeling}, an optically thick point
source has been adopted to fit L1448 IRS 3B data.  First, to verify
that the point source should be optically thick to imply a radial
variation of $\beta$, we tested the case of a point source with the
same $\beta$ to that of its envelope.  As expected, a point source
with the same $\beta$ as the envelope requires a smaller $\beta$
to fit the data (Fig. \ref{fig_ptfluxbeta}).  The dashed contours
in Figure \ref{fig_ptfluxbeta} are 80\%, 60\%, and 40\% of the
peak value in the likelihood distribution of the same models in
Figure \ref{fig_likelihood} and the dotted contours are
80\%, 60\%, and 40\% of the likelihood peak in the new models with
a point source having the same $\beta$ to the envelope. 
A parameter space of $p : 1.9-2.4$ and $\beta : 0.4-0.9$ with
the other parameter ranges the same as the optically thick point source
models, except $R_{in}$ which was fixed at 10 AU, has been searched.  
In addition to the smaller $\beta$, it
is noteworthy that there is no ``good'' fit.  The
``best fit'' gives $\chi_\nu^2 \sim 11$, which is much
worse than the case of the optically thick point source case
($\chi_\nu^2 \sim 8.7$).  This is expected as there is no good way
to well fit the two wavelength data simultaneously without a variable
$\beta$ along radius.  Note that the differences between the two
wavelength amplitudes are only sensitive to $\beta$.  Since we
assume a constant $\beta$ for the point source and the envelope in
the new model, the differences between the two wavelength amplitudes
along radius can be caused only by the optically thick emission
due to the density increase of the inner envelope.  
As the new model is worse than the optically
thick point source model, this test also implies that the optically
thick emission, purely due to the density increase of the inner
envelope of L1448 IRS 3B, is not significant enough to explain the
$\beta$ variation in the data.

\begin{figure}
\begin{center}
\includegraphics[angle=270,scale=0.50]{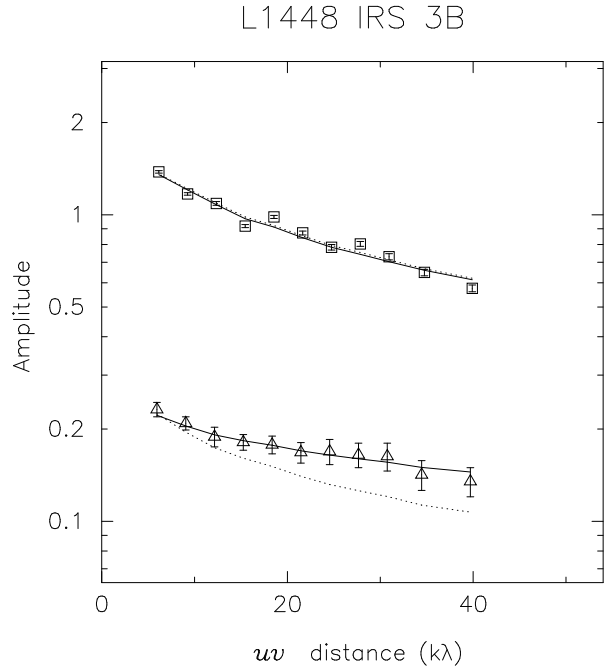}
\end{center}
\caption{Examples of fitting models to emphasize a radial dependence
of $\beta$. The solid lines are the best fit model with $\beta$ as
a function of radius ($\chi_\nu^2 \sim 7.1$) and the dotted lines
present an example of fitting models with a constant $\beta$.
Parameter sets for the best fit model (solid line): $p=2.6$, $M_T=2.20$ \Msol,
$\beta_{out}=1.7$, $R_\beta=400$ AU, $R_{in}=10$ AU, $R_{out}=4500$ AU,
$F_{pt}=0.0$ Jy, $T_0=100$ K at $R_0=10$ AU and for the other one of
a constant $\beta$ (dotted line): $p=2.5$, $M_T=2.80$ \Msol, $\beta=1.0$, 
$R_{in}=10$ AU, $R_{out}=4500$ AU, $F_{pt}=0.0$ Jy, $T_0=100$ K at $R_0=10$ AU.
Note that the data points are the same as in Fig. \ref{fig_uvamp} and the
error bars are statistical standard errors. No absolute flux calibration
uncertainties are shown. 
\label{fig_varBT}}
\end{figure}

Similarly, better (probably more ``realistic'', a sharper temperature
gradient in inner regions) temperature distributions such as of
\citet{looney2003} and \citet{chiang2008} cannot fit the
data either.  We tested simulated temperature
distributions similar to those studies and verified that they do
not provide radially variable differences between the two wavelengths.
The dotted line in Figure \ref{fig_varBT} is an example of fitting
models with the better temperature distribution but with a constant
$\beta$ over radius.  As shown, it does not produce the variable
amplitude differences with radius between the two wavelengths.
It is understandable since the inner regions are hotter resulting in
a valid Rayleigh-Jeans approximation, i.e. no slope
change between the two wavelengths due to temperature variation.

Finally, we construct a model to simulate the variable $\beta$ as a
function of radius, based on grain growth. 
A point source of L1448 IRS 3B seems to be weaker than $20$ mJy 
if it existed, according to \citet{looney2003}, whose data went to
$\sim 400$ \klambda\ at \tmm.  Therefore, modeling with an optically thick
point source is not the best way, although it provides
a relatively ``good'' fit for our intermediate angular resolution data.  
For this reason, we do not consider a point source in the following model.  

We assume grain growth by gas accretion onto grain surfaces.  Grains
can grow by gas accretion and coagulation and can be destroyed or
denuded by grain-grain collisions and heating mechanisms such as
cosmic rays, central protostellar radiation, and bipolar outflow
shock waves \citep[e.g.,][]{draine1985}.  To address grain growth
fully, these growth and destruction mechanisms may need
to be taken into account together.  However, we presume only grain
growth by gas accretion without considering any destruction mechanisms 
for simplicity.  Coagulation might contribute significantly in
the dense envelopes but its efficiency is uncertain
\citep[e.g.,][]{flower2005}.  Grain growth
by coagulation requires relative grain motion, which can be
introduced by various mechanisms.  Relative velocities caused by thermal
movement, ambipolar diffusion, and radiation pressure lead to grain
coagulation rather than grain shattering; the velocities are smaller
than the critical velocities, which are the upper limits of velocity
for grain coagulation depending on grain properties such as size,
composition, and shape.  The critical velocities have been studied
theoretically \citep[e.g.,][]{chokshi1993} and experimentally
\citep[e.g.,][]{blum2000,poppe2000}.  However, the velocity is too
small to consider coagulation as an efficient mechanism for
grain growth \citep{draine1985}.  Alternatively, hydrodynamically or 
magneto-hydrodynamically
induced turbulence \citep[e.g.,][]{yan2004} could bring a faster
relative velocity of grains.  However, it depends on the maximum
velocity at the incident scale, which is highly uncertain, and it
may also lead to grain destruction due to high velocities.
In addition, even when considering the fastest relative velocity
of grains for coagulation (the critical velocity), coagulation may
not be as efficient as gas accretion \citep{flower2005}.

The grain growth rate by gas accretion has a relationship with density
and temperature distributions, $da/dt \propto
w \rho \propto T^{1/2} \rho$, where $a$ and $w$ indicate a grain 
size and a colliding gas velocity \citep[][p 208]{spitzer1978ppim}.
Note that although we assume only grain growth by gas accretion, 
grain growth rate by coagulation has a similar relationship with
the density and relative velocity of grains instead of gas density
and velocity.  Overall, this formulation is arguably valid for
a general description of grain growth, in a well-mixed gas 
and dust region.
In addition to the grain growth rate, we simply
assume that $\beta$ is inversely proportional to the maximum grain
size \citep{draine2006}. Therefore, after some time period, $\beta(r)$
is inversely proportional to the product of the density distribution
and the square root of the temperature distribution,
\begin{displaymath}
\beta(r) = \left\{ \begin{array}{ll}
\beta_{out} (r/R_\beta)^p\, (T_d(R_\beta)/T_d(r))^{1/2} &  \textrm{where }
r \leq R_\beta \\
\beta_{out} & \textrm{where } r > R_\beta.
\end{array} \right.
\end{displaymath}
We fix $\beta_{out} = 1.7$ \citep[e.g,][]{draine2006} and instead 
introduce $R_\beta$ for an adjustment of the radial dependence.
In addition, we allow  the temperature distribution to change along 
$\beta(r)$.  However, $T_d(r) = T_0 (R_0/r)^{2/(4+\beta(r))}$ is not a
monotonic function, i.e., presumably not realistic.  
Therefore, we design a temperature distribution smoothly changing 
from a case of $\beta = 0$ to a case of $\beta = \beta_{out}$ around
$R_\beta$,
\begin{equation}
T_d(r) = \frac{W_1(r)\, T_1(r) + W_2(r)\, T_2(r)}{W_1(r) + W_2(r)},
\end{equation}
where $W_1(r)=R_\beta/r$, $W_2(r)=r/R_\beta$, 
$T_1(r)=T_0 (R_0/r)^{2/4}$, and
$T_2(r)=T_0 (R_0/r)^{2/(4+\beta_{out})}$.
We recognize that the temperature distribution might not be the
best one corresponding to the variable $\beta$.  However, we point
out that the temperature distribution mainly changes the flux density
profiles, not the differences between flux densities of the two
wavelengths (Fig. \ref{fig_varBT}).  Therefore, the modeling here
focusing on the variable $\beta$, which is implied for the variable
differences of the flux densities along radius, is not sensitive
to the temperature distribution.  We searched a parameter space of
$p$, $M_T$, $R_{out}$, and $R_\beta$ with the other fixed parameters 
($\beta_{out} = 1.7$, $R_{in}=10$ AU, $F_{pt}=0.0$ Jy, $T_0=100$ K at 
$R_0=10$ AU) as listed in Table \ref{tab_rbeta}.  Figure \ref{fig_rbeta}
shows the result, a likelihood distribution on $p$ vs. $R_\beta$.
The $p$ and $R_\beta$ are most likely to be about 2.6 and 400 AU,
respectively.  The parameter set of the best fit model ($\chi_\nu^2
\sim 7.1$) is $p=2.6$, $M_T=2.20$ \Msol, $R_\beta=400$ AU, and
$R_{out}=4500$ AU and the averages weighted by the likelihood are
$p=2.59$, $M_T=2.51$ \Msol, $R_\beta=420$ AU, and $R_{out}=5900$
AU.  The best fit model is plotted in Figure
\ref{fig_varBT} overlaid with the observational data.

In this model, the best fit suggests an
envelope that is mostly ``interstellar medium grains'' (small grains with $\beta\sim1.7$), with grain growth at the very center,
$R_\beta~\lesssim~400$ AU, which is approximately the smallest structure sensitivity of these observations.
It is important to note that this is not equivalent to models of an ``interstellar medium grain'' envelope with a point source of a smaller $\beta$ value, 
as those models do not fit (Fig. \ref{fig_ptfluxbeta}), and in addition, such a bright point source
at \tmm\ is not consistent with the results of \citet{looney2003}.

The $p$ value ($\sim 2.6$) is larger than the value ($\sim 2.1$) obtained
in \S\ \ref{sec_modeling} assuming an optically thick point source.
This is understandable because applying a point source itself causes
a density gradient, as mentioned in \S\ \ref{sec_modeling} for L1157.
Actually, this $p$ value is more consistent with the results of
\citet{looney2003} using larger {\it uv} coverage data and a
higher angular resolution at \tmm.  Based on the facts that the
data of L1448 IRS 3B do not have a point source feature and that
this model has a smaller $\chi_\nu^2 \approx 7.1$, we argue that
the larger $p$ from this model is more reliable.  

To understand the large difference between $p$ values of L1448 IRS 3B and
the other two sources, we focus on the differences of the apparent
properties.  While L1448 IRS 2 and L1157 are isolated and have a
very large bipolar outflow ($\sim 5 '$), L1448 IRS 3B is in a
``binary system'' and its bipolar outflow is not so extended 
\citep[e.g.,][]{kwon2006}.  These facts imply that the density
distribution could be steeper in binary and/or younger (based on
the kinematic time scales of bipolar outflows) YSOs such as L1448
IRS 3B.  \citet{looney2003}, who have carried out {\it uv} modeling
towards 6 sources, have also reported relatively steeper density
distributions for bright YSOs of ``binary systems'' such as NGC
1333 IRAS 4B and L1448 IRS 3B.
However, density indexes larger than $2$ are somewhat puzzling, since they
indicate expansion rather than collapse, i.e., the thermal pressure
gradient exceeds the gravitational force.  However, we might be
able to connect this aspect to their binarity, in which the outer
envelope is affected by the companion, or their youngness, in which
the envelope is affected by the bipolar outflow momentum.  Detailed
theoretical studies are needed to understand this.

The $R_\beta$ value indicates an outer limit where grain growth 
mainly occurs. 
According to \citet{spitzer1978ppim}, the grain growth rate by
gas accretion in the diffuse interstellar medium ($T=80$ K, $n_H=20$ 
cm$^{-3}$) is given by,
\begin{equation}
\frac{da}{dt} = 2\times10^{-12}\, \xi_a\, \Big(\frac{T}{80 \textrm{ K}}\,
\frac{1}{\mu}\Big)^{1/2}\, \Big(\frac{n_H}{20 \textrm{ cm}^{-3}}\Big)\, 
\frac{\textrm{mm}}{\textrm{year}},
\end{equation}
assuming a typical dielectric grain density and a cosmic composition
gas.  The $\xi_a$ is a sticking probability, and the $\mu$ is the mean
gas particle weight.  
Although grain growth in dense regions such as the central regions of
Class 0 YSO envelopes could be different, it is applicable as discussed
before.
Simply compensating for our temperature ($\sim 40$ K), 
the mean gas particle weight increase 
(two-atomic molecular gas rather than atomic gas), and density ($n_{H}
\sim 10^9$ cm$^{-3}$ at 200 AU), we can obtain $da/dt =
5 \times 10^{-5} \xi_a$ (mm/year).  When accepting 
$\xi_a=1$,\footnote{Although \citet{spitzer1978ppim} assumed $\xi_a=0.1$ 
for the diffuse interstellar medium, $\xi_a=1$ is arguably a better
assumption for the cold and dense inner envelope regions
\citep[e.g.,][]{flower2005}.}
this implies that a time scale of $10^4$ years, comparable to the kinematic
time scales of bipolar outflows of Class 0 YSOs
\citep[e.g.,][]{bachiller2001}, can result in about mm-size grains. 
Although grain growth could also  occur
in previous stages, it is much more efficient in the higher densities
of the Class 0 stage.  Another interesting point is that
less massive (i.e., less bright) and less steep density distribution
envelopes such as those of L1448 IRS 2 and L1157 would have smaller
radial regions for the grain growth
within the same time scale. Then, in such sources, the variation of
$\beta$ may not be distinct nor distinguishable from a point source,
as shown in \S\ \ref{sec_vis_comp}. 

We interpreted the radial dependence of $\beta$ based on grain
growth above. However, there could be another effect, grain
segregation. \citet{ciolek1996} showed that magnetic fields in
protostellar cores reduce abundances of small grains in the cores
by a factor of its initial mass-to-magnetic field flux ratio.  In
other words, a stronger magnetic field with respect to the mass of a
core causes more effective segregation.  Although this segregation
occurs while the ambipolar diffusion appears, before dynamical
collapse, the signature footprint could remain in the envelopes of
Class 0 YSOs. On the other hand, although this effect would be minor to
the features we have discussed because the segregation is effective 
to relatively small grains ($a \lesssim 10^{-4}$ cm), it is noteworthy 
that it would set the initial grain distribution of Class 0 YSO envelopes
for more efficient growth in the central region.

\section{Conclusion}

We carried out interferometric observations towards three
Class 0 YSOs (L1448 IRS 2, L1448 IRS 3, and L1157) at \omm\ and 2.7
mm continuum using CARMA.  The continuum at these millimeter
wavelengths is mainly thermal dust emission of their envelopes. Our
observations have been designed particularly to cover comparable
{\it uv} ranges at the two wavelengths, which allowed us to tackle dust
grain opacity spectral indexes ($\beta$) of Class 0 YSOs, using
unprecedented compact configuration and high image fidelity.
Through simultaneous modeling of the two wavelength visibilities
as well as comparisons of the images and visibilities for the first time, 
we found not only the $\beta$ of Class 0 YSOs but also
its radial dependence. In addition, we addressed
the single power-law density index $p$ of Class 0 YSO envelopes.

1. We found that the dust opacity spectral index $\beta$ of the
earliest YSOs, so-called Class 0, is around 1. This implies that
dust grains have significantly grown already at the earliest stage.

2. We obtained the power-law density index $p$ of $\sim 1.8$, $\sim
2.6$, and $\sim 1.7$ for L1448 IRS 2, L1448 IRS 3B, and L1157,
respectively.  Although we did not attempt to constrain star formation
theories, we pointed out the difference between that of L1448 IRS
3B and those of the other two.  Based on different properties of
L1448 IRS 3B from the other two sources, we suggested that ``binary
system'' YSOs and/or younger YSOs in terms of kinematic time scales
of their bipolar outflows would have steeper density distributions.

3. We found radial dependences of $\beta$. In particular, the dependence is
distinct in L1448 IRS 3B.  We verified it by models employing $\beta$ as a
function of radius. In addition, we discussed that the grain growth causing
the dependence can be achieved in a time scale of $10^4$ years, corresponding
to the kinematic time scale of bipolar outflows of Class 0 YSOs.

\acknowledgments
First of all, we thank the CARMA staffs for their dedicated work 
to commission and operate CARMA.
W. K. thanks M. W. Kunz, T. Ch. Mouschovias, and C. F. Gammie for
helpful discussions and comments.  In addition, we thank anonymous
referee for valuable comments to improve this paper.
W. K. and L. W. L. acknowledge
support from NASA Origins Grant No. NNG06GE41G.  L. G. M. acknowledges
support from NASA Origins Grant No. NNG06GE16G.
Support for CARMA construction was derived from the states of Illinois,
California, and Maryland, the Gordon and Betty Moore Foundation, the
Eileen and Kenneth Norris Foundation, the Caltech Associates, and the
National Science Foundation.  Ongoing CARMA development and operations
are supported by the National Science Foundation under cooperative 
agreement AST-0540459, and by the CARMA partner universities.

Facilities: \facility{CARMA}

%\appendix

\bibliographystyle{apj}
\bibliography{apj-jour,kwon_dustcont}

\clearpage

\begin{table}
\begin{center}
\caption{Targets and Observations \label{tab_obs}}
\begin{tabular}{lllllll}
\tableline\tableline
Source & $\alpha$ (J2000.0) & $\delta$ (J2000.0) & & & & \\
Wavelength & Date & Flux cal. & Gain cal. & Flux & Array & Beam size (PA)\tablenotemark{a} \\
\tableline 
L1448 IRS 2 & 03 25 22.346 & +30 45 13.30 & & & & \\
\tableline 
1.3 mm & 2007 Aug. 21 & Uranus & 3C84 & 4.0 & E & 
$5\farcs3 \times 4\farcs4$ ($- 72 \arcdeg$) \\ 
  &  &  & 0237+288 & 1.2 &   &  \\ 
2.7 mm & 2006 Sep. 02 & Uranus & 0237+288 & 1.6 & Comm.\tablenotemark{b} & 
$4\farcs8 \times 4\farcs3$ ($- 74 \arcdeg$) \\ 
 & 2006 Sep. 12 & Uranus & 0237+288 & 1.6 & Comm. &  \\
\tableline
L1448 IRS 3 & 03 25 36.339 & +30 45 14.94 & & & & \\
\tableline 
1.3 mm & 2007 Aug. 19 & Uranus & 3C84 & 3.9 & E & 
$5\farcs0 \times 4\farcs3$ ($71 \arcdeg$) \\
 &  &  & 0237+288 & 1.2 &  &  \\
2.7 mm & 2006 Dec. 03 & Uranus & 0237+288 & 1.88 & Comm. & 
$5\farcs0 \times 4\farcs5$ ($43 \arcdeg$) \\ 
\tableline
L1157 & 20 39 06.200 & +68 02 15.90 & & & & \\
\tableline 
1.3 mm & 2007 Aug. 20 & MWC349\tablenotemark{c} & 1927+739 & 0.95 & E & 
$4\farcs6 \times 3\farcs8$ ($24 \arcdeg$) \\ 
2.7 mm & 2007 Jul. 12 & Mars & 1927+739 & 1.6 & D & 
$7\farcs0 \times 5\farcs6$ ($7 \arcdeg$) \\ 
\tableline 
\end{tabular}
\tablenotetext{a}{The synthesized beam in the case of natural weighting.
}
\tablenotetext{b}{An array configuration for commissioning tasks, 
similar to D.  Note that only part of the array was available in some cases.}
\tablenotetext{c}{The flux is assumed as 1.8 Jy, based on periodic CARMA flux 
calibrator measurements.}
\end{center}
\end{table}

%\clearpage

\begin{table}
\begin{center}
\caption{Weighting and Tapering Schemes and Final Synthesized Beams \label{tab_beam}}
\begin{tabular}{cccccc}
\tableline\tableline
Source & Frequency \tablenotemark{a} & Weighting & Tapering & 
       Beam Size (PA)\tablenotemark{c} & Beam Ratio \\
       & (GHz) & (Robust factor)\tablenotemark{b} & & & (1 mm / 3 mm) \\
\tableline
L1448 IRS 2 & 228.60 & 0.8 & & $4\farcs986 \times 4\farcs168$ ($-78.91 \arcdeg$) & \\
            & 112.94 & natural & & $4\farcs826 \times 4\farcs277$ ($-74.06 \arcdeg$) & 1.007 \\
\tableline
L1448 IRS 3 & 228.60 & natural & & $5\farcs049 \times 4\farcs299$ ($70.87 \arcdeg$) & \\
            & 112.84 & 1.1 & & $4\farcs951 \times 4\farcs412$ ($43.29 \arcdeg$) & 0.994 \\
\tableline
L1157  & 228.60 & natural & $5\farcs6 \times 6\farcs1$ & $5\farcs597 \times 5\farcs026$ ($-10.95 \arcdeg$) & \\
            & 113.00 & 0.0 & & $5\farcs644 \times 5\farcs015$ ($-3.850 \arcdeg$) & 0.994 \\
\tableline
\end{tabular}
\tablenotetext{a}{The frequencies used for $\beta$ calculation. Refer to 
eq. (\ref{eq_beta}).}
\tablenotetext{b}{Briggs' robust weighting factor \citep{briggs1995}.}
\tablenotetext{c}{Beam size uncertainties are order of $0.1\arcsec$, and 
the values shown are to illustrate the beam size ratios.}
\end{center}
\end{table}

%\clearpage

\begin{table}
\begin{center}
\caption{$\beta$ values of the sources \label{tab_beta}}
\begin{tabular}{cccc|ccc}
\tableline\tableline
 & \multicolumn{2}{c}{Fluxes (Jy)} & & \multicolumn{3}{c}{$\beta$ maps} \\
Sources & 1.3 mm & 2.7 mm & $\beta$ & Minimum & Maximum & Average  \\
\tableline
L1448 IRS 2 & $0.20$ & $0.025$ & $0.95$ & $0.70$ & $1.6$ & $1.0$ \\
L1448 IRS 3 &       &       &      &         &       & $0.60$ \\
L1448 IRS 3A & $0.090$ & $0.012$ & $0.85$ & $0.32$ & $1.7$ & $0.90$ \\
L1448 IRS 3B & $1.0$ & $0.19$ & $0.35$ & $-0.14$\tablenotemark{a} & $1.7$ & $0.53$ \\
L1448 IRS 3C & $0.15$ & $0.026$ & $0.48$ & $0.12$ & $2.1$ & $0.59$ \\
L1157       & $0.29$ & $0.050$ & $0.49$ & $-0.008$\tablenotemark{a} & $1.3$ & $0.47$ \\
\tableline
\end{tabular}
\tablenotetext{a}{The negative $\beta$ values are due to a bias introduced in
deconvolution.}
\end{center}
\end{table}
 
%\clearpage

\begin{table}
\begin{center}
\begin{small}
\caption{Model parameter sets for the three sources \label{tab_param}}
\begin{tabular}{lcccccccc}
\tableline\tableline
\multicolumn{1}{c}{Targets} & & $p$ & $\beta$ & $M_T$ & $R_{in}$ & $R_{out}$ & $F_{pt}$\tablenotemark{a} & $T_0$\tablenotemark{b} \\
        &   &   &         & (M$_\sun$) & (AU) & (AU) & (Jy) & (K) \\
\tableline
L1448 IRS 2 & $\Delta$\tablenotemark{c} & $1.5-2.0$ & $0.5-1.5$ & $1.00-2.00$ & $10-30$ & $4000-6000$ & $0$ & $100$ \\
  & $\delta$\tablenotemark{d} & $0.1$ & $0.1$ & $0.05$ & $10$ & $500$ & $-$\tablenotemark{e} & $-$ \\
  & best\tablenotemark{f} & 1.8 & 0.9 & 1.35 & 10 & 5500 & 0 & 100 \\
  & mean\tablenotemark{g} & 1.79 & 0.88 & 1.36 & 20 & 5300 & 0 & 100 \\
\tableline
L1448 IRS 3B & $\Delta$ & $1.8-2.4$ & $0.7-1.3$ & $3.25-4.35$ & $10-20$ & $4000-7000$ & $0.06-0.12$ & $100$ \\
  & $\delta$ & $0.1$ & $0.1$ & $0.05$ & $10$ & $500$ & $0.01$ & $-$ \\
  & best & 2.2 & 1.1 & 3.25 & 10 & 6500 & 0.120 & 100 \\
  & mean & 2.14 & 0.96 & 3.68  & 14   & 6300 & 0.099 & 100 \\
\tableline
L1157 & $\Delta$ & $1.5-2.0$ & $0.5-1.5$ & $0.30-1.00$ & $10-30$ & $1000-3000$ & $0.000-0.035$ & $100$ \\
  & $\delta$ & $0.1$ & $0.1$ & $0.05$ & $10$ & $500$ &  0.005 & $-$ \\
  & best & 1.8 & 0.8 & 0.55 & 30 & 2000 & 0.015 & 100 \\
  & mean & 1.73 & 0.91 & 0.59  & 20   & 2300 & 0.019 & 100 \\
\tableline
  & $\Delta$\tablenotemark{h} & $1.5-2.0$ & $0.5-1.5$ & $0.30-1.00$ & $10-30$ & $1000-3000$ & $0.015-0.025$ & $100$ \\
  & mean\tablenotemark{h} & 1.72 & 0.91 & 0.59  & 20   & 2300 & 0.020 & 100 \\
\tableline
\end{tabular}
\tablenotetext{a}{A central point source flux at \tmm. Here the point sources
are assumed as optically thick indicating $\beta=0$.}
\tablenotetext{b}{Temperature at $R_0 = 10$ AU}
\tablenotetext{c}{Parameter range searched}
\tablenotetext{d}{Parameter steps}
\tablenotetext{e}{Fixed parameter}
\tablenotetext{f}{Best fitting parameter set with the smallest $\chi_\nu^2$}
\tablenotetext{g}{Mean of parameters weighted by the likelihood, 
$\textrm{exp}(-\chi_\nu^2/2)$}
\tablenotetext{h}{These two lines present the cases of models with a limited
point source flux range. Refer to the text.}
\end{small}
\end{center}
\end{table}

%\clearpage
 
\begin{table}
\begin{center}
\caption{Model parameter sets with $\beta$ as a function of radius 
for L1448 IRS 3B. 
\label{tab_rbeta}}
\begin{tabular}{lcccccccc}
\tableline\tableline
\multicolumn{1}{c}{Targets} & & $p$ & $R_\beta$ & $M_T$ & $R_{in}$ & $R_{out}$ & $F_{pt}$ & $T_0$ \\
        &   &   &  (AU)   & (M$_\sun$) & (AU) & (AU) & (Jy) & (K) \\
\tableline
L1448 IRS 3B  & $\Delta$ & $2.4-2.8$ & $250-550$ & $2.20-3.20$ & $10$ & $4000-7000$ & $0.00$ & $100$ \\
  & $\delta$ & $0.1$ & $50$ & $0.05$ & $-$ & $500$ & $-$ & $-$ \\
  & best & 2.6 & 400 & 2.20 & 10 & 4500 & 0.000 & 100 \\
  & mean & 2.59 & 420 & 2.51  & 10 & 5900 & 0.000 & 100 \\
\tableline
\end{tabular}
\end{center}
\end{table}

%\clearpage

%\clearpage

%\clearpage

%\clearpage

\end{document}